\documentclass[letterpaper, 10pt]{article}

\usepackage[margin=2.5cm]{geometry}
\usepackage{gensymb}
\usepackage{booktabs}
\usepackage{graphicx}
\usepackage{caption}
\usepackage{subcaption}
\usepackage{float}
\usepackage{amsmath}
\usepackage{algorithmicx}
\usepackage{subfiles}
\usepackage{multirow}
\usepackage{url}

\usepackage{textcomp} 
\usepackage{placeins}
\usepackage{nameref}

\usepackage{color}
\usepackage{xcolor}
\usepackage{colortbl}
\usepackage{soul}

\usepackage{tikz}
\usetikzlibrary{calc}
\usepackage{flushend}
\usepackage{bigstrut}
\usepackage{makecell}

\usepackage[pdfencoding=auto, pdfusetitle,hidelinks]{hyperref}
\hypersetup{
    colorlinks,
    linkcolor={black!50!black},
    citecolor={black!50!black},
    urlcolor={black!80!black}
}

\makeatletter
\let\@fnsymbol\@arabic
\makeatother

\newcommand\saystop{\mbox{``\textit{stop}'' }} 

\hypersetup{
  pdftitle={Personal space of autonomous car's passengers sitting in the driver's seat},
  pdfauthor={Eleonore Ferrier-Barbut, Dominique Vaufreydaz, Jean-Alix David, J\'er\^ome Lussereau, Anne Spalanzani},
  pdfkeywords={Virtual Reality, Autonomous car, Personal space, Human Vehicle Interaction, Proxemics},
}

\title{Personal space of autonomous car's passengers sitting in the driver's seat}

\author{Eleonore Ferrier-Barbut$^1$, Dominique Vaufreydaz$^2$, Jean-Alix David$^1$,\\J\'er\^ome Lussereau$^1$, Anne Spalanzani$^1$}

\date{
\textsuperscript{1}~Univ. Grenoble Alpes, Inria, Grenoble INP, 38000 Grenoble, France\\
\textsuperscript{2}~Univ. Grenoble Alpes, CNRS, Inria, Grenoble INP, LIG, 38000 Grenoble, France\\
\vspace{0.5em}\textit{Author Version}
}

\begin{document}

\maketitle
\thispagestyle{plain}
\pagestyle{plain}

\begin{abstract}
This article deals with the specific context of an autonomous car navigating in an urban center within a shared space between pedestrians and cars. 
The driver delegates the control to the autonomous system while remaining seated in the driver's seat. The proposed study aims at giving a first insight into the definition of human perception of space applied to vehicles by testing the existence of a personal space around the car.
It aims at measuring proxemic information about the driver's comfort zone in such conditions.
Proxemics, or human perception of space, has been largely explored when applied to humans or to robots, leading to the concept of personal space, but poorly when applied to vehicles. 
In this article, we highlight the existence and the characteristics of a zone of comfort around the car which is not correlated to the risk of a collision between the car and other road users. 
Our experiment includes 19 volunteers using a virtual reality headset to look at 30 scenarios filmed in 360$\degree$ from the point of view of a passenger sitting in the driver's seat of an autonomous car.
They were asked to say \saystop when they felt discomfort visualizing the scenarios.

As said, the scenarios voluntarily avoid collision effect as we do not want to measure fear but discomfort.
The scenarios involve one or three pedestrians walking past the car at different distances from the wings of the car, relative to the direction of motion of the car, on both sides. The car is either static or moving straight forward at different speeds.
The results indicate the existence of a comfort zone around the car in which intrusion causes discomfort.
The size of the comfort zone is sensitive neither to the side of the car where the pedestrian passes nor to the number of pedestrians. 
In contrast, the feeling of discomfort is relative to the car's motion (static or moving).
Another outcome from this study is an illustration of the usage of first person 360$\degree$ video and a virtual reality headset to evaluate feelings of a passenger within an autonomous car.
\end{abstract}

\section{INTRODUCTION}

\subsection{Proxemics theory}
There is, surrounding individuals, a dynamic zone which they actively defend, affirms Hall \cite{hall1966hidden}, the introducer of proxemics. It is an area in which any intrusion causes discomfort, as proved by Hayduk \cite{hayduk1981permeability}. 
In another study \cite{hayduk1994personal}, it is shown that this space is different from a territory, given the latter's frontiers are defined and stable in time, whereas personal space is a ''momentary spatial preference``. 

The existence of this space is either allocated to physical and social protection objectives or to navigation objectives. The Dosey-Meilsels Protection Theory \cite{dosey1969personal}, assigns personal space to protection objectives from a physical attack towards one's body or social attack towards one's social image, whereas \cite{gerin2008characteristics} attributes it to navigation's planning objectives.
Hence, \cite{d2017invisible,patane2016disentangling,lloyd2009space} demonstrated that there are two different  types  of personal spaces. They are linked to different high-order representations of the body. 
First, the interpersonal space rather depends on social factors and consequently varies considerably with one's culture, emotional state or relationship with the observed individual. It is a space of security and protection.
Then, the peripersonal space is a space around the body used to interact with close objects. It is a working space, in which manipulations occur, used to anticipate and navigate.
It varies depending on accessibility, environment, speed or direction.
The dimensions of these two spaces were proved to be independent in \cite{patane2016disentangling}. A tool held in hand by a subject for example, changing the individual's accessibility to the space surrounding him, modifies the perceived dimensions of the peripersonal space around him, but doesn't modify those of the interpersonal space surrounding him. 
This statement is also defended in \cite{higuchi2006action}, arguing that representation of peripersonal space is modified when an external object is incorporated in the body schema, to cover areas immediately surrounding the object. 

\subsection{Proxemics and cars}
Recent researches by McLaughling \cite{mclaughlin2016understanding}, and more specifically, by Lubbe \cite{lubbe2014pedestrian,lubbe2015drivers}, suggest that this concept of proxemics and particularly of a comfort zone surrounding the individual can be equally applied to cars.
These studies argue that there is a space around the car defined by comfort boundaries, which is the space kept by the driver between his car and other road users on his trajectory, as a result of a constant calculation to avoid collision.
It is a space actively defended by the driver, who pushes the brakes whenever other road users intrude or are about to intrude this comfort zone.
Its dimensions are measured, in \cite{lubbe2014pedestrian,lubbe2015drivers} with respectively one hundred and eight, and sixty-two volunteers.
These experiments involve a car and a pedestrian crossing over in front of the car.
Subjects were driving the car, and were surprised by a pedestrian crossing in front of them.
At the moment when they push the brakes, measures are taken of the Time To Collision (TTC), lateral distance between the center of the pedestrian and center of the driving lane, and longitudinal distance between the center of the pedestrian and the front of the car.
It is demonstrated, in \cite{lubbe2015drivers}, that the dimensions of this space depend significantly on the pedestrian's speed.
The higher is the pedestrian's speed, the larger are the comfort zone's dimensions.
These experiments test the dimensions of the comfort space around the car as a function of a risk of collision, as opposed to the peripersonal space mentioned above.
The idea that pedestrian's speed and comfort boundaries around the car are highly correlated is also supported in \cite{ren2016analysis}.

\subsection{Comfort in autonomous cars}
In this paper, we investigate the case of an autonomous car navigating in an urban center within a shared space between pedestrians and cars.
Such spaces, as imagined by town planners like Ben Hamilton-Baillie and Paul Boston for instance, start to appear in cities.
In our scenario, the driver delegates the control to the autonomous system while remaining seated in the driver's seat.
The car navigates at slow speed among other cars and pedestrians within a shared space.
The slow speed allows communication between the autonomous car and surrounding pedestrians (\cite{RasouliKT17,Schneemann2016}) and thus increase safety.

In this study,
we postulate that there is a zone around the car, which persists to exist independently from the driver's calculation of risk of collision and in which any intrusion causes discomfort. 
The Social robotics field has proposed a great amount of work on human-aware navigation, or socially-aware navigation that integrates social spaces in the navigation decision process \cite{Kirby2009, Pandey2010, Sisbot2007, Rios11}.
The question of human acceptance of autonomous cars behavior starts to be seen in the field of autonomous cars \cite{rothenbucher2016ghost, basu2017you, hauslschmid2017supportingtrust}.

We propose that the driven individual inside the car does not forget its sense of a personal space around him because of the protection afforded by the car's bodywork.
Rather, he sees its dimensions widen by both the extended frontiers of the space he occupies on the road and his augmented accessibility of the surrounding space depending on his vehicle's direction and speed.
Our scenario differs from former experiment \cite{lubbe2015drivers}.
Pedestrians walk tangentially to the car without apparent risk of collision.  

{In the case of the autonomous car and when the passenger sitting in the driver's seat have delegated control to the autonomous driving system}, it is predicted that awareness about a zone around the car that has to be defended persists.
We estimate that, even in the absence of a prediction related to navigation, the passenger on board an autonomous car will feel discomfort when a pedestrian walks too close to the car.
Consciousness of this space may depend on many factors such as willing to defend one's property on the car, relationship with the person coming close to the car, one's emotional state, in comparison with the personal space in the case of the individual described in \cite{hall1966hidden}.

\begin{figure*}[ht!]
\centering
\includegraphics[width=\textwidth]{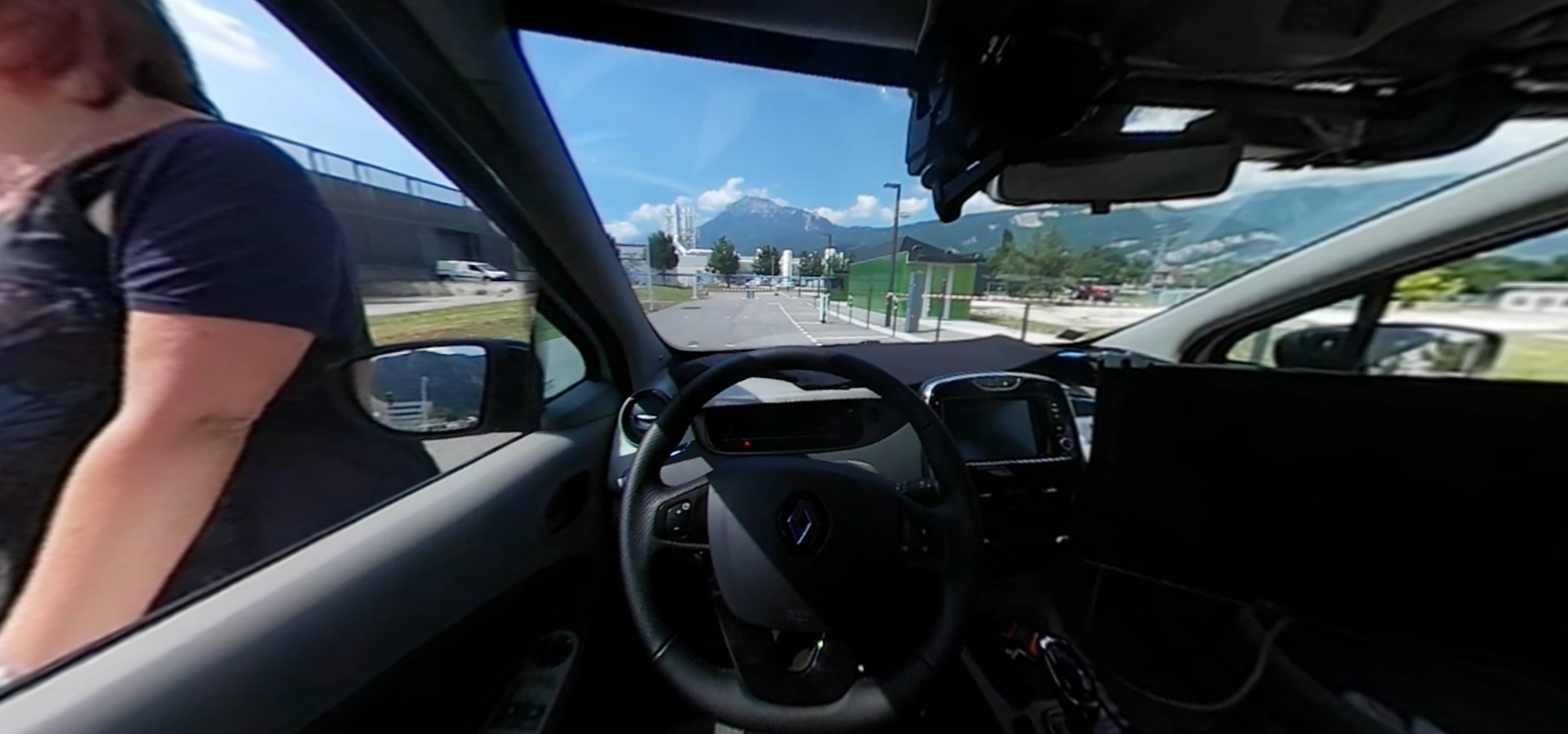}
\caption{Example of a pedestrian passing by the car. The screen shoot is extracted from the 360$\degree$ video as seen by subjects in the virtual reality headset. On this image, the pedestrian is on the left at 0.2m from the car. On top, one can see the camera mounted on the ceiling. On bottom right, the screen of the control computer of the autonomous car (off to avoid distraction).}
\vspace{-0.3cm}
\label{fig:360video}
\end{figure*}

Many factors enter in the process of evaluating the existence of a comfort zone around the car that can be seen as a personal space. One can cite the risk of collision, the speed of the car, the speed of the pedestrian, etc.

Our experiment aims at testing whether a driven individual inside an autonomous car still has, in the absence, thus, of control over the car, a mental representation of a comfort space surrounding the car. In other words, we tested if the subject's representation of the zone to defend around him extended its boundaries from his body to the car's bodywork when being inside an autonomous car. In order to do so, we selected the factors which were likely to provoke the subject's feeling of discomfort characteristic of an intrusion in the comfort zone, such as Hayduk defines it in \cite{hayduk1981permeability}. 
The factors that we controlled were the ones which were likely to test something different from what we were testing, such as the risk of collision and the speed of the pedestrian tested in \cite{lubbe2014pedestrian,lubbe2015drivers}. They were all the factors linked with taking decisions related to the navigation of the car, directly dependent on the control over the car.
Therefore, we eliminated the subject's control over the car by virtually boarding them into an autonomous car using a virtual reality (VR) headset and by fixing the car's direction.
Regarding the pedestrian, still within a posture to avoid testing the calculation of a risk of collision, we fixed the pedestrian's direction and his speed, and we eliminated any kind of communication from the pedestrian. Pedestrian was walking past the car in the opposite direction, without colliding with the car (figures \ref{fig:360video} and \ref{fig:arc}).

As regards the factors of our experiment, we first decided to test the effect of the car's speed. If the subject felt discomfort related to the approach of the pedestrian in a car over which he had no control, was it only when the car was moving or also when it remained static? Then, we decided to test the influence of the side on which the pedestrian walked past, to determine whether the feeling of discomfort arose only when the pedestrian was close to the pedestrian's body, or to the whole car's bodywork. To have thresholds from the comfort zone's dimensions, we also tested the influence of the distance from the pedestrian to the car relatively to the motion of the car (see figure \ref{fig:arc}.). We experimented if there is a threshold at which the subjects significantly started to feel discomfort. To finish with, we tested whether the number of pedestrians influence significantly the number of reported feeling of discomfort by the subjects.

In summary, we tested the feeling of discomfort of nineteen subjects virtually boarded onto an autonomous car as a function of four parameters: the car's speed, the proximity from the pedestrian to the car, the side of the car on which the pedestrian is walking past and the number of pedestrians walking past.
The discomfort is evaluated using a procedure inspired by \cite{hayduk1994personal}: subjects express their discomfort saying \saystop while watching the scenarios.
We hypothesized that feeling of discomfort would be more frequently observed in scenarios in which pedestrians were closer to the car, and in which the car was going faster.
We expected that the number of pedestrians would have an effect on the measured dependent variable \saystop answers; a larger number of pedestrians would generate a significantly higher number of \saystop answers.
Also, that the number of \saystop answers would be significantly larger for scenarios in which the pedestrian is walking past the car on the left, compared to when it is walking past on the right, because it is closer to the subject's body.

\section{METHOD}

The experimental method relies on recordings made by a 360$\degree$ camera involving an autonomous car and 1 or 3 pedestrians walking past it.
Recent works investigated the effect of 360$\degree$ videos and headset on affective labeling by users compare to a standard monitor condition \cite{McKeown2017}.
They showed that there is no significant effect of this modality.
Using this technology, we expect to observe the same neutrality while benefiting from immersive performance of the device.

To assess the personal space around the car, subjects viewed 360$\degree$ videos using a Virtual Reality headset and evaluate their feeling about the proximity of the pedestrian to the car at several distances and speeds. The experimental room was a quiet room, in which nothing else than the experiment was going on. 

In the next sections, we will present the autonomous car used in our experiment, the recording of the videos and then depict the test with subjects.

\subsection{Autonomous car}

For the experiments, a Renault Zoe electric car has been equipped with a Velodyne HDL64 on the top, 3 Ibeo Lux LiDARs on the front and 1 on the back, Xsens GPS and IMU providing vehicle velocity and orientation, a stereo camera and 2 IDS cameras. Data from LiDARS are synchronized using ROS and fused into occupation grids \cite{rummelhard2017}. The perception system is implemented on a PC in the trunk of the car equipped with a NVidia Titan X GPU, while the previously described automation process has been integrated in the vehicle. The control of the vehicle is implemented in a wired kit using BUS-CAN and direct control on the power steering.
For the sake of the experiments, a dedicated portion of road has been designed and equipped with sensors and security requirements (PTL platform of IRT Nanoelec).

\begin{figure}[h!]
\centering
\includegraphics[width=0.6\linewidth]{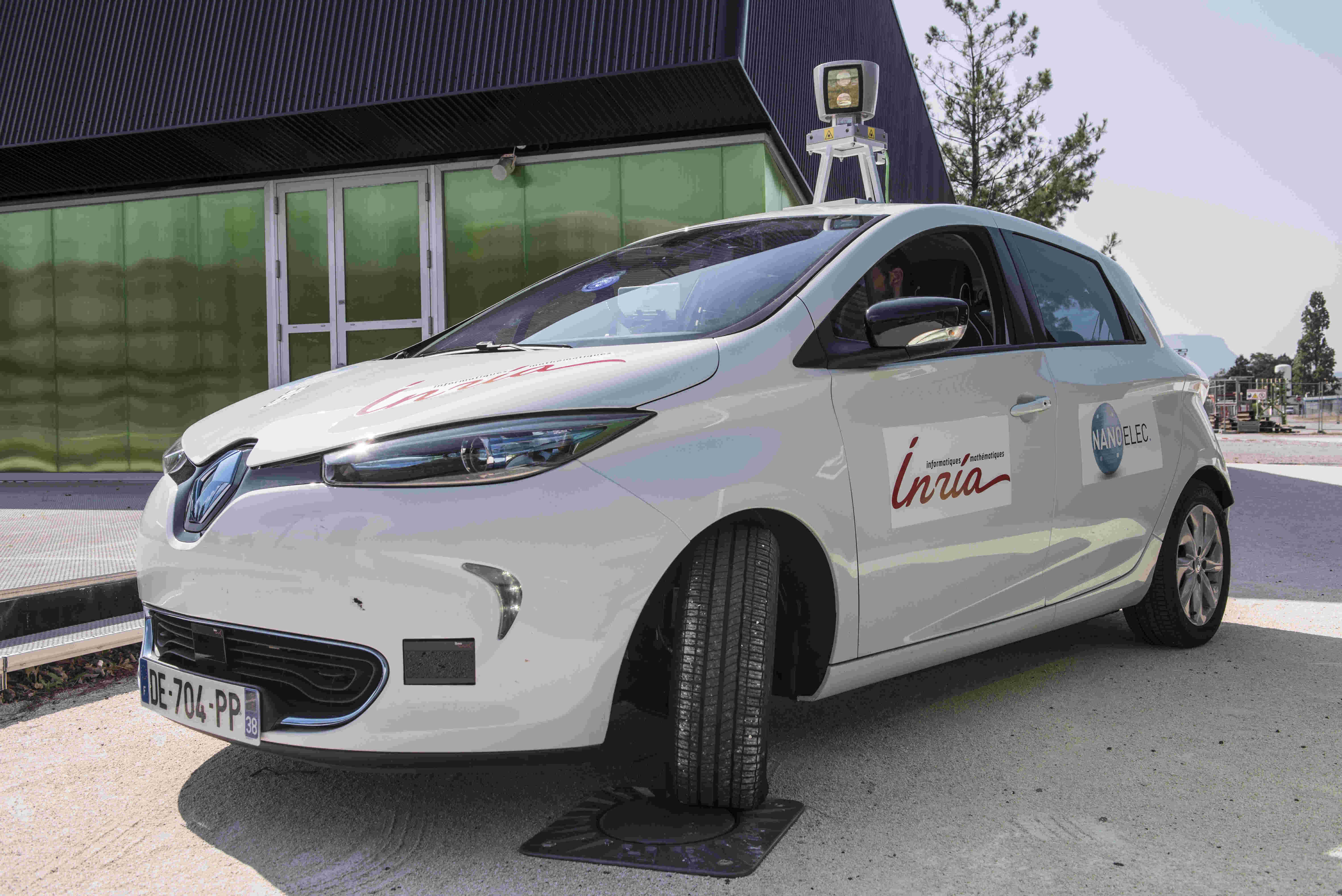}
\caption{The autonomous electric car used in our experiment is a Renault Zoe. It is equipped with a Velodyne HDL64, 4 Ibeo Lux LiDARs, Xsens GPS and IMU, a stereo and 2 IDS cameras.}
\vspace{-0.3cm}
\label{fig:Zoe}
\end{figure}

\begin{figure}[t]
\centering
\includegraphics[width=0.4\linewidth]{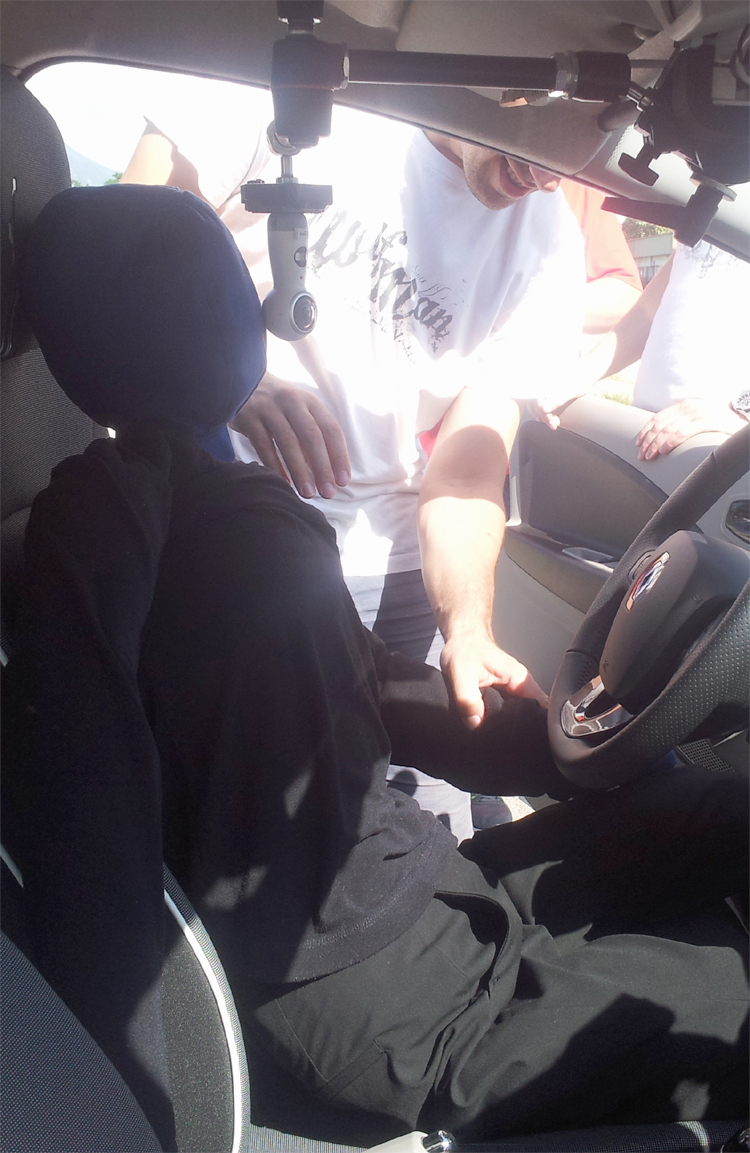}
\caption{Installation of the soft dummy and the 360$\degree$ camera within the autonomous car.}
\vspace{-0.3cm}
\label{fig:Inside}
\end{figure} 

~\\
~\\
~\\
\subsection{Recordings}

To record the video sequences presented to the participants, we used an autonomous electric car from our laboratory.
This car is a white small city car, 1.73~m width, 4.08~m long.
In order to create our scenarios, the car was equipped with the following items (see Figure \ref{fig:Inside}):
\begin{itemize}
\item 1 camera \textit{Samsung Gear 360} fixed to the ceiling at the level of the virtual driver's eyes
\item a dressed soft dummy was placed in the driver's position.
\end{itemize}

This setup was designed to give the participants the impression that they were sitting in the driver's seat of the autonomous car.
Since the videos were 360$\degree$ (Figure \ref{fig:360video}), the participants could look all around them. 
The soft dummy is used to enhance the embodiment: if people looked down with the VR headset, they will see belly and legs of the dummy.
Contrarily, except if they look strictly behind them, they cannot see the dummy's upper body in the seat.
During the capture, no hand can be seen on the wheel or on the gear lever but while the car is moving, the wheel may turn to reflect small correction on the direction of the car (to keep it in a straight line, regardless of the ground).
The car was driven from the outside by an agent with a remote control who could not be seen from the point of view of the camera.
The speed is limited by the control computer embedded in the car: the car cannot go over the desired speed.

\begin{figure}[t!]
\centering
\frame{\includegraphics[width=0.75\linewidth]{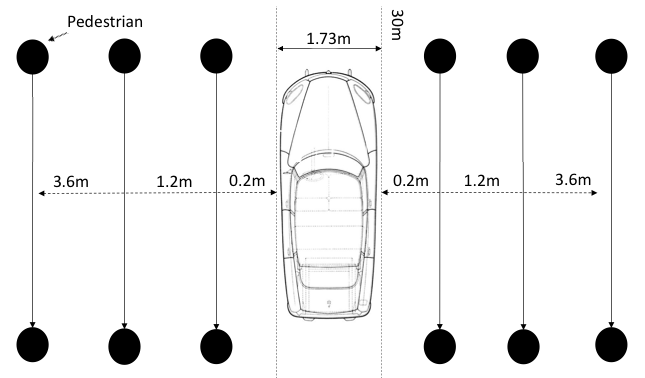}}
\caption{Design of the scenarios. Starting 30~m away, pedestrian(s) pass by the car at each side of the car at 0.2~m, 1.2~m and 3.6~m. The car is a left-hand drive, 1.73~m width, 4.08~m long. No scenario involves a risk of collision.}
\vspace{-0cm}
\label{fig:arc}
\end{figure}

The figure \mbox{\ref{fig:arc}} depicts the scenarios.
Pedestrian starts 30 meters in front of the car and comes toward it following a line at 3 distances on both sides of the car.
He was instructed to walk straight forward at constant speed, looking away in front of him and not inside the autonomous car.
Group of 3 pedestrians repeated the same recording procedure.
They walk side by side using the previous instructions.
They did not interact with each other.
The distance is evaluated from the closest pedestrian of the group.

36 scenarios were actually captured (6 of them were used for trial with every subject, see section \ref{sec:TestEvents}).
The scenarios filmed the autonomous car being either static or driving autonomously straight forward.
The speed of the car varied, it was either 0~km/h (static car), 7.5~km/h or 15~km/h, in a straight line motion. The position from the pedestrian relatively to the car varied from left to right, and its distance from the car was either 0.2~m, 1.2~m or 3.6~m, measured from the wings of the car (see figure \ref{fig:arc}). The number of pedestrians varied between one and three. 
The 36 scenarios present every possible combination of the modalities of the four variables Speed (3 values), Distance (3), Laterality (2), Number of Pedestrians (2).

\subsection{Subjects}

19 volunteers submitted to the experience, all personnel or master students of our research center, some very used to new technologies, others less. They were fifteen men and four women. They all had a driver's license.

\begin{figure}[h!]
\centering
\includegraphics[width=0.45\linewidth]{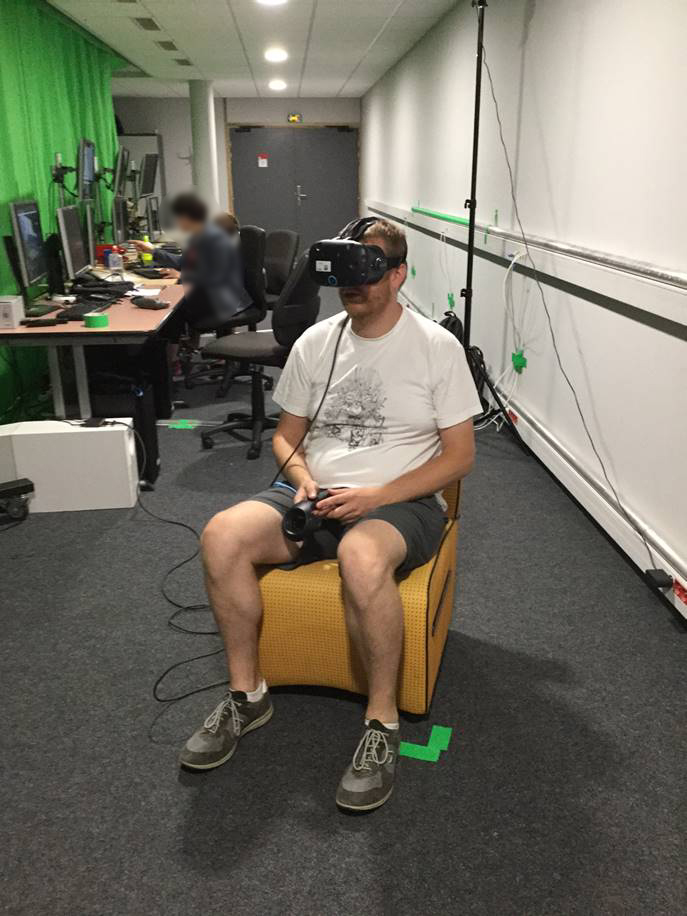}
\caption{Subject during the experiment. He is wearing the virtual reality headset while sitting in an armless seat. An experimenter is reporting \saystop answers when the subject expresses discomfort with the pedestrian(s) passing by the car.}
\vspace{-0.5cm}
\label{fig:Casque}
\end{figure} 

\subsection{Test Events}
\label{sec:TestEvents}

The figure \ref{fig:Casque} presents a typical test.
The VR headset is a \textit{HTC Vive\copyright} with a horizontal and a vertical fields of view respectively of 100$\degree$ and 110$\degree$. 
Head movements are captures using embedded inertial sensors and an external laser positioning system.
Subjects were sitting within an experimental room in a seat without armrest as if they were in the car.
We have envisioned to carry out experiments directly into the car to further enhance the immersive effect of the videos.
Unfortunately, this was not possible because of the headset tracking system which is not usable inside a vehicle.

A total of 30 responses were collected for each of the 19 subjects, representing 570 responses.
Of these, about one-fifth is from women (120) and four-fifths are from men (450).
If we check different conditions, lateralization and velocity, the same number of responses was given for each possible value.
For instance, 150 responses for men for each possible speed.
Identically, 40 responses were gathered from women.
However, for the number of pedestrians and distances, the number of response collected varies.
We have fewer answers for 1.2~m (114 instead of 228) and for 3 pedestrians (228 instead of 342).
This is explained by the choice of videos showed to familiarize subjects with the experimental setup, guarantying that they were comfortable with the VR headset and that they could see everything well.
Indeed, 6 scenarios, involving three pedestrians passing by the car at a distance of 1.2~m at every speed, were used in this trial session.
We chose not to test these 6 scenarios because the number of pedestrians was not the most important variable we wanted to test, and by keeping the 12 other scenarios where 3 pedestrians could be seen we had enough data to test the influence of this variable, at every speed and on each side.

The 30 remaining scenarios were shown in three randomized orders corresponding to three blocks: blocks 1 and 2 were shown to six participants, block 3 to seven participants.
Each block contains all 30 scenarios.
The participants were instructed to say \saystop when the proximity from the pedestrian to the car in the scene made them feel discomfort, a procedure we copied from Hayduk's experiments \cite{hayduk1994personal} called the stop distance procedure. 
The \saystop answers were written down by an external reporter for each subject with their matching scenario as ``1'' (respectively as ``0'' when nothing was said).
The percentage of \saystop answers was measured for each condition. 

\begin{table}[h!]
\centering
\resizebox{0.6\linewidth}{!}{%
    \begin{tabular}{|c|c|c|c|c|}
\cline{3-5}    \multicolumn{1}{c}{} &       & \textbf{Women} & \textbf{Men} & \textbf{Total} \bigstrut\\
    \hline
    \cellcolor[RGB]{242,242,242}  & \cellcolor[RGB]{242,242,242} \textbf{0.2m} & \cellcolor[RGB]{242,242,242} 70.83\% & \cellcolor[RGB]{242,242,242} 42.78\% & \cellcolor[RGB]{242,242,242} 48.68\% \bigstrut\\
\cline{2-5}   \cellcolor[RGB]{242,242,242}       & \cellcolor[RGB]{242,242,242} \textbf{1.2m} & \cellcolor[RGB]{242,242,242} 16.67\% & \cellcolor[RGB]{242,242,242} 4.44\% & \cellcolor[RGB]{242,242,242} 7.02\% \bigstrut\\
\cline{2-5}  \multirow{-3}[6]{*}{\cellcolor[RGB]{242,242,242}\textbf{Distance}}        & \cellcolor[RGB]{242,242,242} \textbf{3.6m} & \cellcolor[RGB]{242,242,242} 0.00\% & \cellcolor[RGB]{242,242,242} 1.11\% & \cellcolor[RGB]{242,242,242} 0.88\% \bigstrut\\
    \hline
    \multirow{2}[4]{*}{\textbf{\begin{tabular}[c]{@{}c@{}}Number of\\ pedestrians\end{tabular}}} & \textbf{1} & 30.56\% & 18.89\% & 21.35\% \bigstrut\\
\cline{2-5}          & \textbf{3} & 33.33\% & 17.78\% & 21.05\% \bigstrut\\
    \hline
    \cellcolor[RGB]{242,242,242}\ & \cellcolor[RGB]{242,242,242} \textbf{static} & \cellcolor[RGB]{242,242,242} 7.50\% & \cellcolor[RGB]{242,242,242} 4.67\% & \cellcolor[RGB]{242,242,242} 5.26\% \bigstrut\\
\cline{2-5}  \cellcolor[RGB]{242,242,242}        & \cellcolor[RGB]{242,242,242} \textbf{7.5} & \cellcolor[RGB]{242,242,242} 45.00\% & \cellcolor[RGB]{242,242,242} 26.67\% & \cellcolor[RGB]{242,242,242} 30.53\% \bigstrut\\
\cline{2-5} \multirow{-3}[6]{*}{\cellcolor[RGB]{242,242,242} \textbf{Speed}}         & \cellcolor[RGB]{242,242,242} \textbf{15} & \cellcolor[RGB]{242,242,242} 42.50\% & \cellcolor[RGB]{242,242,242} 24.00\% & \cellcolor[RGB]{242,242,242} 27.89\% \bigstrut\\
    \hline
    \multirow{2}[4]{*}{\textbf{Laterality}} & \textbf{Left} & 28.33\% & 21.78\% & 23.16\% \bigstrut\\
\cline{2-5}          & \textbf{Right} & 35.00\% & 15.11\% & 19.30\% \bigstrut\\
    \hline
    \multicolumn{2}{|c|}{\cellcolor[RGB]{242,242,242} \textbf{All conditions}} & \cellcolor[RGB]{242,242,242} 31.67\% & \cellcolor[RGB]{242,242,242} 18.44\% & \cellcolor[RGB]{242,242,242} 21.23\% \bigstrut\\
    \hline
    \end{tabular}%
}
\caption{Positive response rate for each variable, by gender and in total.}
\label{tab:gender}
\end{table}

\section{RESULTS}

Results of our experiment have been statistical analyzed through a Cochran's Q Test \cite{cochran1950comparison}.
Influence of every factor (Speed, Distance, Laterality and Number of pedestrians) on the measured dependent variable was analyzed using a binary logistic regression, on which an Anova was performed.
All the results presented in this section, unless explicitly stated differently, are significant.

\begin{table}[!ht]
\parbox[t][][t]{0.48\linewidth}{
\centering
\vspace{1em}
\resizebox{0.85\linewidth}{!}{
\begin{tabular}{|c|c|c|c|}
\cline{2-4}\multicolumn{1}{r|}{} & \textbf{0.2 m} & \textbf{1.3 m} & \textbf{3.6 m} \bigstrut\\
\hline
\textbf{left} & 52.63\% & 7.02\% & 0.00\% \bigstrut\\
\hline
\textbf{right} & 44.74\% & 7.02\% & 1.75\% \bigstrut\\
\hline
\end{tabular}%
}
\caption{Percentage of positive answers for each side at every distance (all speeds, 1 or 3 pedestrians)}
\label{figdisgd}
}
\parbox[t][][t]{0.48\linewidth}{
\centering
\vspace{1em}
\resizebox{0.9\linewidth}{!}{
\begin{tabular}{|c|c|c|c|}
\cline{2-4}\multicolumn{1}{r|}{} & \textbf{0.2 m} & \textbf{1.3 m} & \textbf{3.6 m} \bigstrut\\
\hline
\textbf{static} & 11.84\% & 2.63\% & 0.00\% \bigstrut\\
\hline
\textbf{7.5 km/h} & 68.42\% & 13.16\% & 1.32\% \bigstrut\\
\hline
\textbf{15 km/h} & 65.79\% & 5.26\% & 1.32\% \bigstrut\\
\hline
\end{tabular}%
}
\caption{Percentage of positive answers for each distance at every speed (both sides, 1 or 3 pedestrians)}
\label{figdistvit}
}
\hspace{1cm}

\parbox[t][][t]{0.48\linewidth}{
\centering
\vspace{1em}
\vspace{1em}
\setlength\intextsep{0pt}
\resizebox{0.85\linewidth}{!}{
\begin{tabular}{|c|c|c|c|}
\cline{2-4}\multicolumn{1}{r|}{} & \textbf{static} & \textbf{7.5 km/h} & \textbf{15 km/h} \bigstrut\\
\hline
\textbf{left} & 4.21\% & 35.79\% & 26.32\% \bigstrut\\
\hline
\textbf{right} & 6.32\% & 25.26\% & 29.00\% \bigstrut\\
\hline
\end{tabular}%
}
\caption{Percentage of positive answers for each side at every speed (all distances, 1 or 3 pedestrians)}
\label{figvitgd}
}
\hspace{0.02\linewidth}
\parbox[t][][t]{0.48\linewidth}{
\centering
\vspace{1em}
\vspace{1em}
\resizebox{0.85\linewidth}{!}{
\begin{tabular}{|c|c|c|}
\cline{2-3}\multicolumn{1}{r|}{} & \textbf{0.2 m} & \textbf{3.6 m} \bigstrut\\
\hline
\textbf{1 pedestrian} & 4.21\% & 35.79\% \bigstrut\\
\hline
\textbf{3 pedestrians} & 6.32\% & 25.26\% \bigstrut\\
\hline
\end{tabular}%
}
\caption{Percentage of positive answers for each number of pedestrians at 0.2~m and 3.6~m (both sides, all speeds)}
\label{figpieton}
}
\end{table} 

\subsection{Effect of the distance to the car}

Pedestrian's distance to the car is significant. Results of table \ref{tab:gender} showed 48.68\% of subjects saying \saystop when the pedestrian is walking 0.2~m from the wing of the car, compared to only 7.02\% when the pedestrian is 1.2~m far from the wing of the car, and $<$1\% when he is 3.6~m away.
Table \ref{figdisgd} shows the percentage of positive answers at every distance and for left and right sides. 
There is no significant difference between both sides.
0.2~m remains the most uncomfortable distance.
Table \ref{figdistvit} depicts positive rates at every distance and several speeds and show no significant effect of the distance.

Looking at these results, it is likely to say that the comfort zone is not affected by the position of the person within the car.
Indeed, the distance effect is the same for both side.
The personal space of the driver seems to extend to the body of the car.

\subsection{Effect of the speed of the car}

Speed alone was also found to have a significant effect. 27.89\% of subjects said \saystop at 15~km/h, 30.53\% at 7.5~km/h and 5.26\% only when the car remained static (table \ref{tab:gender}).
We conducted further statistical analysis on the effect of distance and laterality to the speed (see table \mbox{\ref{figvitgd}}).
It appears that there is no significant difference between the number of \saystop answers for 0.2~m of distance from the pedestrian to the car and 3.6~m of distance when the car is static, either when the pedestrian is passing by the car on the right or on the left.
There is however a significant difference (Cochran's Q Test $p<.05$) on both sides between the number of \saystop answers at a distance of 0.2~m and of 3.6~m when the car is moving at 15~km/h, showing an interaction effect between speed and distance.
We didn't test the significance of this difference at 7.5km/h because the results at 15km/h and at 7.5km/h are not significantly different ($p>.05$). 
Results on both sides of the test are presented in table \mbox{\ref{figvitgd}}.

\subsection{Effects of the number of pedestrians and laterality}

Contrary to what we expected, neither the number of pedestrians (see table \ref{figpieton}) nor the laterality (cf table \ref{tab:gender}) did show any significant effect. 
19.30\% of subjects said \saystop when the pedestrian was walking on the left from the car, and 23.2\% when the pedestrian was walking on the right from the car, 21.3\% when one pedestrian was walking past the car, and 22.5\% with three pedestrians. 

The absence of significant effect from the laterality of the pedestrian supports the idea that the mental representation of the comfort zone of an individual, when inside a car, extends its frontiers to the car's bodywork, letting him sensitive to approaches towards the car's boundaries.

In our experiment, pedestrians were asked to stay neutral, go straight forward without looking into the car.
Identically for the group of 3 pedestrians.
In this condition, they walk side by side.
This neutral behavior may explain why the number of pedestrians does not affect significantly the results.

\subsection{Gender differences}

The distribution between men and women in our corpus is unbalanced. The women's responses represent a total of 120 out of 570 responses in total (450 for men).
However, it is interesting to see if there is a difference between men's and women's personal space projection around the autonomous car.
We are aware that these results specifically require further experimentation to demonstrate their relevance as they are not statically significant.
We therefore present here only trends that have yet to be confirmed.

In table \ref{tab:gender}, one can find the percentage of positive answers for each tested variable for each gender (2 central columns).
The first information read in these results is that there seems to have a greater discomfort for women than for men.
This fact is true for all scenarios except for a distance of 3.6~m where men seem more uncomfortable (1.11\% for men compared to 0\% for women)
Nevertheless, this difference is not significant.

\section{DISCUSSION}

The results indicate that there is an awareness of the closeness from pedestrian to the car in which subjects are virtually boarded, revealed by the fact that subjects significantly indicated a feeling of discomfort when this proximity was too great.
When there is no risk of collision and the passenger from the car is not supposed to take any decision related to the driving, there still is an awareness of a space around the car in which intrusion causes discomfort. That is revealed by the fact that 48.6\% of subjects declared a feeling of discomfort when a pedestrian walked past the car at a distance of 0.2~m, whereas only 7\% did when the pedestrian was 1.2~m from the car, and $<$1\% only did when the pedestrian was at a distance of 3.6~m, differences which proved to be significant ($p<.05$).

The fact that the car is in motion appears to be a major factor influencing the consciousness of a comfort zone around the vehicle which has to be defended, causing discomfort when intruded. Indeed, the Distance effect is not significant when the car is static. This might indicate that the subjects virtually boarded onto the car felt protected by the car's bodywork when it was static. This feeling of security inverses when the car is in motion, state in which the Distance effect proved to be very significant. This indicates that the subjects were  conscious of their increased accessibility of the space on the road, and also that they appropriate the car's frontiers. Indeed, the notable increase of the number of \saystop answers when the car was in motion and the pedestrian at a distance of 0.2~m probably indicates a will to protect the pedestrian walking past, related to the awareness of being in his personal zone. It is a different form of discomfort than that felt when one's personal space boundaries are crossed. In the car's case, when the car is in motion, the feeling of discomfort is caused by the awareness of being a potential intruder in other's personal space, what's more physically much stronger. These results are consistent with the idea of a social space around the car, of protection from physical or social attacks, and not only of planning for navigation. Surprisingly, the percentages of \saystop answers were less significant when the car was moving at 15~km/h compared to when it was moving at 7.5~km/h, which might be due to the fact that the two speeds are hardly differentiable in the videos, but we cannot make any statement. 

\clearpage
Looking at results of our study, some conclusions about our variables can be drawn:
\begin{itemize}
\item The most important distance is 0.2~m.
\item The number of pedestrians and lateralization have no significant effect.
\item The difference between a stationary and a moving car is important, but there is no significant difference between 7.5km/h and 15km/h.
\item These results are the same for both genders. However, one can note that a distance of 1.2~m appears to have a greater effect for women (as already said, this result is just a trend in our study as we have less women's answers.).
\end{itemize}

These results suggest that the car's bodywork is incorporated in the body schema of passengers like in former work from Higuchi \& al. \cite{higuchi2006action}.
This is one of our expectation but surprisingly this personal space extension is not lateralized.
At both side of the car, the effect of speed and distance are the same.
These results with an autonomous car are coherent with former studies about proxemics \cite{lubbe2014pedestrian,lubbe2015drivers,mclaughlin2016understanding}.
This consistency of our results is an indicator of neutrality of the 360$\degree$ modality, as Mc Keown \& al. in their study \cite{McKeown2017}.
One may say that the 360$\degree$ condition can be used to test people reaction in autonomous driving condition.
Nevertheless, more experiments must be conducted to definitely prove it.

\section{CONCLUSION}

Investigation of the comfort zone around the car, in the autonomous car's case, indicates that it is influenced by different factors than those which have an effect on individual's personal space. Indeed, the existence of a comfort zone around the car suggested by the results is directly conditioned by the physical form of the car and by its abilities. The results are consistent with the idea that the car is both a protection for the passengers aboard the car and a threat for the pedestrians passing near the car. Indeed, when the car is static, the fact that the subjects don't significantly feel discomfort related to the proximity from the pedestrian to the car may indicate that the will to defend a zone around the car is mainly related to wanting to protect other road users rather than themselves, protected by the car. Equally, the absence of significant effect of the laterality of the pedestrian supports the idea that the individual inside the car is not protecting himself but the car he is in, probably from hurting others and from being damaged. Nonetheless, the results allow us to say the individual's sensitivity is extended to the car's bodywork.

For further work, the results encourage to extend these experiments with more people, balancing women and men, ages and driver/non-driver subjects.
This will allow us to confirm drawn hypotheses, and to validate those on gender differences.
It would also be interesting to change the subject point of view to include front and rear passenger views as well.
These results also encourage conducting further research on the factors influencing the dimensions of the comfort zone around the car, in order to have their more precise measures (between 0.2 and 1.2~m as it is the questioned area). Recording biometric data such as heart rate would probably add interesting information.  We must question the causes of one's consciousness of the car's physical boundaries, and discomfort when these are approached by others.

Another underlying outcome of our study is the usage of a 360$\degree$ camera and a virtual reality helmet in experiments about the feelings of passengers within an autonomous car.
Consistency of our results is an indicator of neutrality of the 360$\degree$ modality in this context.
It also makes this study reproducible by other research teams to confirm/refute/refine these results in other contexts or culture.
This can be done either by using our recordings directly with other subjects, or by reproducing our experimental setup.

\section{ACKNOWLEDGMENT}
The experiment data were recorded using hardware provided by Amiqual4Home (ANR-11-EQPX-0002), Kinovis (ANR-11-EQPX-0024) and with the autonomous car and PTL platform from the IRT Nanoelec, founded by the French program Investissement d'Avenir (ANR-10-AIRT-05). Thanks to Laurence Boissieux for helping us in this project and to the PIA Ademe CAMPUS Project which financed the internship.


\bibliographystyle{unsrt}
\bibliography{Biblio.bib}

\end{document}